\documentclass[12pt,a4paper]{article}
\pdfoutput=1
\usepackage[utf8]{inputenc}
\usepackage[T2A,T1]{fontenc}
\usepackage[english]{babel}
\usepackage{amssymb,amsfonts,amsmath,mathtext,cite,enumerate,float} %
\usepackage[dvips,pdftex]{graphicx}
\usepackage{geometry} %
\usepackage{ifpdf}
\usepackage{hyperref}
\title{Residual resistance simulation of an air spark gap switch.}

\author{ V.~V.~Tikhomirov, S.E. Siahlo\thanks{E-mail:svetaju@mail.ru}}

\begin{document}
         \maketitle
         \begin{center}
                  Research Institute for Nuclear Problems, Belarusian State University,\\
Bobruiskaya 11, 220030 Minsk, Belarus
         \end{center}

\begin{abstract}

The numerical simulation of an air spark gap has been carried within two theoretical models. The kinetic one \cite{1} allowed us to calculate time dependencies for the residual resistance (0.2 - 0.4 Ohm for our selection of a circuit parameters), the spark gap channel width, the electron number density, the mobility, the conductivity, the ionization degree, the magnetic field in the discharge channel, the channel inductance and the electron drift velocity. Simulating a real circuit and taking into account a spark gap residual resistance demonstrates good agreement of both models with the experimental data, while that without taking into account this resistance overestimates the maximal current in the circuit by approximately  5$\%$.

\end{abstract}

\section{Introduction}

The high-voltage spark gap switches are essential elements of the circuits containing magneto-cumulative generators, capacitive and inductive energy storages, electro-explosive circuit breakers etc. Very frequently spark gaps affect sufficiently the operation of the circuits and one cannot satisfactorily describe main elements functioning in the absence of a quantitative model of the processes occurring in the spark gaps. This article is devoted to the comparison of two such models considering a real circuit, where the influence of a high-voltage spark gap results in a decrease of the maximal current for about 5$\%$.

We review in detail the model \cite{1} based on the time dependent method for the electron number density simulation in the air spark gap channel using an anisotropic solution of the kinetic equation \cite{2}. The solution \cite{2} was obtained taking into account vibrational, dissociation, excitation, and ionization losses and is applicable at field strengths $E/p_{0}$ lying between 40 and 450 V/cm-mm Hg. This model was successfully implemented for the experimental data interpretation in studying nanosecond high-voltage spark gaps \cite{1} at the upper region of the the specified range of $E/p_{0}$.

We, on the contrary, apply the model \cite{1} in the lower region of the allowable field strengths $E/p_{0}$ appearing in the spark gap at a voltage of about 15 -20 kV (that corresponds about 50 V/cm-mm Hg). Apart from our main objective to describe experimentally observed influence of the spark gap residual resistance on the functioning of the whole circuit, we simulate various physical parameters of the spark gap channel and compare simulation results with those obtained using known model \cite{3}.

\section{Physical models}

In \cite{1} simulation of all physical processes in a spark gap (dissociation, molecules ionization and excitation, electron recombination etc.) is reduced to the calculation of the electron number density depending on the voltage $U_{s}(t)$ and the field strength $E(t)=U_{s}(t)/d$, where $d$ is the gap length. We assume the field to be uniform along the discharge channel following \cite{4}, where it has been demonstrated that during a short enough time period (comparing to the time of a spark gap breakdown) the field distribution becomes nearly uniform. Also in \cite{4},  based on the numerical simulations of the discharge dynamics in the streamer model within 2D and 1D approximations, it is stated that one can describe the processes in a spark gap channel with a reasonable accuracy within 0D approximation.

 The electron number density is defined by the kinetic model \cite{1} on the assumption of an anisotropic distribution function \cite{2} and can be written as

\begin{equation}
\label{eq1}
n_{e}(t)= n_{e}(0)\exp^{\overline{\gamma} t},
\end{equation}

\noindent where $n_{0}(t)= 1 cm^{-3}$ is the electron initial number density in the air and \cite{1},

\begin{equation}
\label{eq2}
\overline{\gamma}=2\left(\frac{\overline{\nu} _x\Delta\varepsilon}{\varepsilon _i}\right)\left(1+\frac{3}{z}+\frac{3}{z^2}\right)\exp^{-z} ,
\end{equation}

\noindent is the mean growth rate for the free electron number density. Following further  \cite{1} $z=\left(\frac{6 \overline{\nu} _x\Delta\varepsilon}{\xi _0}\right)^{1/2}$, $\overline{\nu} _x\Delta\varepsilon=1.19\cdot\left(\frac{p}{p_0}\right) ^{5/6}E^{1/2} $ J/s, $p$ is the air pressure in a spark gap, $p_0$ is the air pressure at sea level, $\overline{\nu} _x$ is the excitation frequency of the air particles, $\Delta\varepsilon$ is the excitation energy, $\varepsilon _i=2.35\cdot 10^{-18}$ J is the ionization energy. $\xi _0=\frac{e^2 E^2}{2m_e \nu _c}$ is the rate at which electrons are gaining energy in the applied electric field $E$,

\begin{equation}
\label{eq3}
\nu _c =1.11 n_a^{5/6}\sigma _c \left(\frac{eE}{m_e}\right)^{1/2}=6.93\cdot 10^{26}\left(\frac{p}{p_0}\right)^{5/6}E^{1/2}\sigma _c \quad s^{-1}
\end{equation}

\noindent  is the elastic collision frequency of a free electron with the air molecules, $e$ is the electron charge, $m_e$ is the electron mass, $n_a$ is the number density of gas particles, $\sigma _c$ is the cross section  for the inelastic collision of an electron with an air particle and it is gained from the experiment and depends on the air pressure \cite{1}.

Knowing the time dependent electron number density, one can easily find the spark gap resistance

\begin{equation}
\label{eq4}
R(t)=\frac{d}{es_z(t)n_e(t)\mu _e(t)} ,
\end{equation}

\noindent  or the current flowing through it

\begin{equation}
\label{eq5}
I(t)=\sigma(t)E(t)s_z(t)=es_z(t)n_e(t)\mu _e(t)\frac{U_s(t)}{d},
\end{equation}

\noindent   where $\mu_e(t)=\frac{e}{m_e\nu _c}$ is the electron mobility, $s_z(t)=\pi r_z^2(t)$ is the current conductive area, $\sigma (t)=e\mu _e(t)n_e(t)$ is the conductivity and $r_z(t)$ is the discharge channel radius. In most of papers on theoretical calculations one chooses a discharge channel width to be a constant value of about 10 microns \cite{5}. This agrees well with our own simulations. Let us give some additional reasons \cite{6} for our choice of a discharge channel model and its width: 1) if the width of a discharge channel is much greater than its radius, one can consider all the processes in the interelectrode gap to be one-dimensional. 2) The real value of the discharge channel radius defining the channel resistance, generally does not match the value of the luminous column radius which is registered experimentally. 3) The discharge channel radius should be defined by the shortest distance from the channel axis to the space point where the plasma specific conductivity is $e=2.7$ times less than the maximal one. 4) One can consider the discharge channel width  to be approximately constant \cite{4}, this statement was done on the basis of the numerical simulations in the streamer model within 2D approximation. We give below comparative simulation results for some physical quantities in the discharge channel calculated both at its constant and variable width and show that in the model \cite{1} the channel width depends on time, however for to estimate the residual resistance it is quite acceptable to choose it to be constant.

To find a suppositional form of the time dependent width of the current conductive region we use an approximate expression \cite{7}

\begin{equation}
\label{eq6}
r_z(t)=\sqrt{2D(t)t},
\end{equation}

\noindent   where $D(t)$ is the electron diffusion coefficient and one can calculate it by two different formulas \cite{7}

\begin{equation}
\label{eq7}
D(t)=\frac{\omega _t^2(t)}{3\nu _c},
\end{equation}

\noindent    and \cite{3}

\begin{equation}
\label{eq8}
D(t)=\frac{\mu _e(t)p}{en_{mol}},
\end{equation}

\noindent   where $\omega _t=\frac{\nu _c}{n_{mol}\sigma _c}$ is the electron average thermal velocity and $n_{mol}=2.51\cdot 10^{19}$ cm$^{-3}$ is the number density of the air molecules. We set an initial voltage on the spark gap to be $0.0001$ V (if one sets it equal to zero, than there appears a singularity in the system of equations at an initial time moment due to a division by zero), the cathode-anode gap $d=1$ cm, the pressure $p=$ 120 kPa and the cross section $\sigma _c=0.2\cdot 10^{-17}$ m$^2$. There is the only empirical parameter in the model described - the elastic collision frequency of a free electron for which we use the experimental data printed in \cite{1}.

In the second model \cite{3} we use, the air gap is replaced by a current generator and the spark gap is described by a nonlinear second order differential equation of

\begin{eqnarray}
\label{eq9}
y(t)\ddot{y}(t)+\dot{y}(t)(1-\dot{y}(t))+b(1-y(t)-\dot{y}(t))\varphi (ay(t))=0, \nonumber \\
y(t)=\frac{U_s(t)}{U_0},\qquad \varphi(ay(t))=\alpha(E/p)^2,
 \end{eqnarray}

\noindent   which is solved together with the circuit equation. $U_0$ is the breakdown voltage of the spark gap, $\alpha$ is the impact ionization coefficient, we use in our calculations $b=24.472$, $a=15.4\cdot 10^5$ the other coefficients can be found in \cite{3}. Unlike \cite{1} it is impossible to get  time dependencies for the electron number density $n_e(t)$, the discharge channel width $r_z(t)$ etc. in model \cite{3}.

\section{Simulation results}

To compare the predictive capabilities of the models described, we simulated a spark gap as a part of a simple electrical circuit consisting of 19.2 mcF capacitor, 300 mcHn inductance, 0.15 Ohm resistance and initial 15 kV voltage on the capacitor. We compare simulation results with the experimental data on the current in a real circuit with the same parameters.

 \begin{figure}[h]
       \centering
        \includegraphics[width=0.5\linewidth]{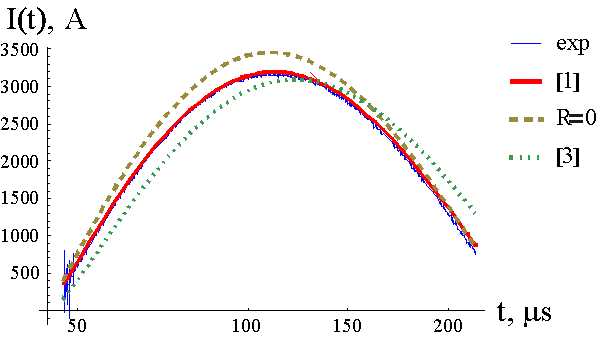}
              \caption{Currents: experimental (solid, blue), simulated by model \cite{1} (thick,red), by model \cite{3} (dashed,green), under the assumption of zero spark gap resistance (dotted,yellow).}

    \end{figure}

There are four current graphs on Fig.1 : the experimental, the simulated by model \cite{1}, by model \cite{3} and under the assumption of zero spark gap resistance. The last graph demonstrates the importance of taking into account a spark gap resistance because there is an overestimation of the maximal current in the circuit by approximately  5$\%$ without it. The currents simulated by models \cite{1,3} are almost identical with the experimental results in the first oscillation half-period in the system which exceeds 200 microseconds.

 \begin{figure}[h]
       \centering
        \includegraphics[width=0.8\linewidth]{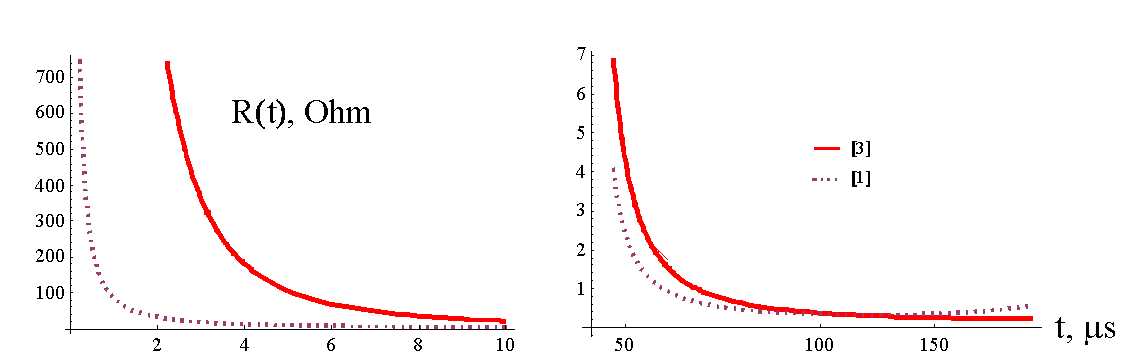}
              \caption{Resistances: simulated by models \cite{1} (thick) and \cite{3} (dashed) on the time intervals (left) from 0 to 10 $\mu$s and (right) from 18 to 200 $\mu$s.}

    \end{figure}

The only drawback is the 8 microsecond current shift in the model \cite{3} with respect to the experimental data. This circumstance is explained by the fact that the model \cite{3} originally is not kinetic, but as is known, kinetic processes dominate in the spark gap in the first microsecond. One can see from Fig.2 that the resistance subsides slower in \cite{3} than in \cite{1} and tends to decrease to a small nonzero value in the model \cite{3}. In the model \cite{1}  the resistance falls to 0.4-0.2 Ohm (depending on the circuit parameters) and then begins to increase, which may indicate spark gap locking. Since we have the ability to compare with a real experiment, we conclude from the above that the model \cite{1} is closer to the real situation than \cite{3}, although they both predict very close values of the resistance and the current.

 \begin{figure}[h]
       \centering
        \includegraphics[width=0.99\linewidth]{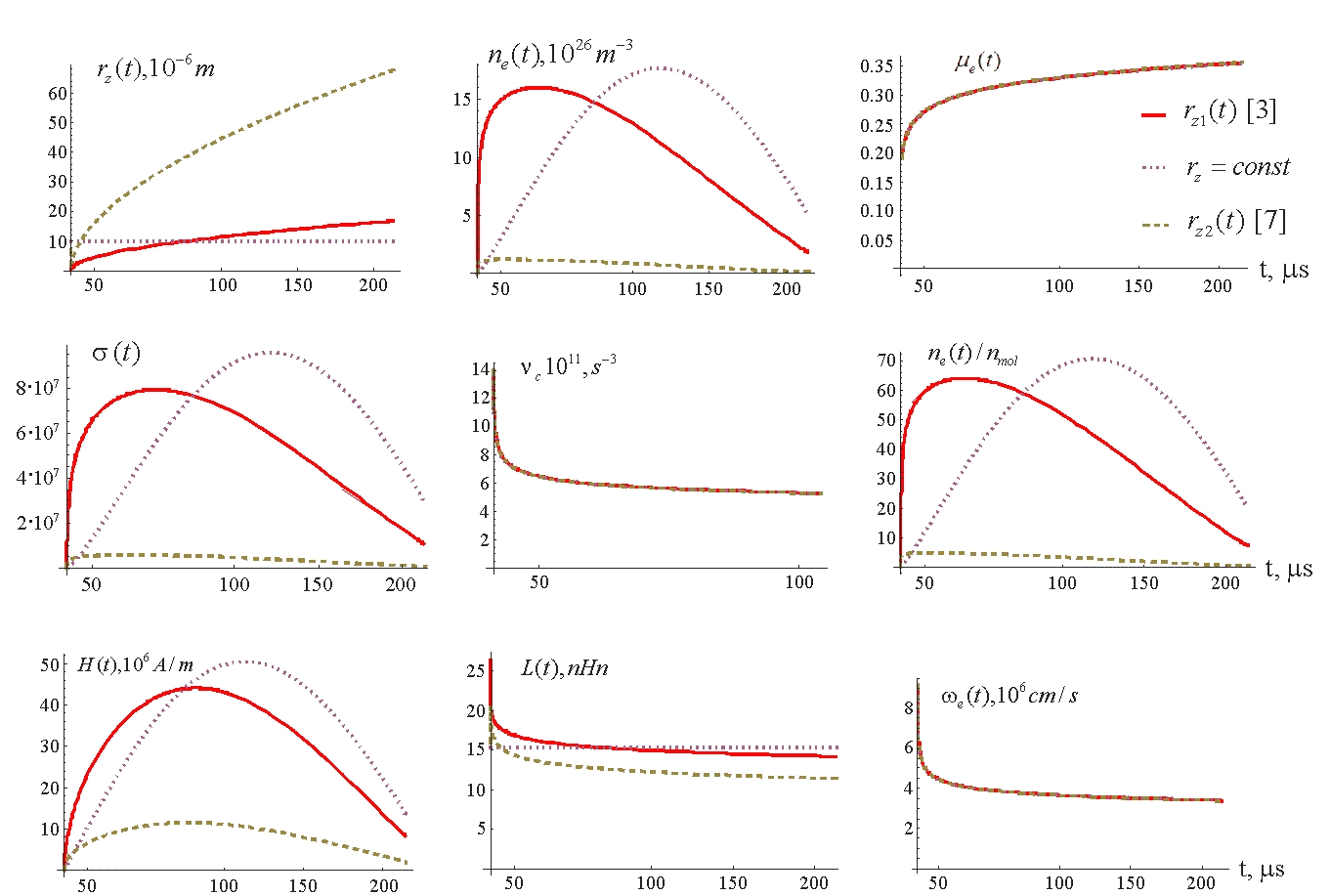}
              \caption{From left to right and from top to bottom: time dependencies for the spark gap channel width $r_z(t)$, the electron number density $n_e(t)$, the mobility $\mu _e(t)$, the conductivity $\sigma (t)$, the collision frequency $\nu _c$, the ionization degree $n_e(t)/n_{mol}$, the magnetic field $H(t)$, the channel inductance $L(t)$ and the electron drift velocity $\omega _e(t)$ simulated by the model \cite{1}.}
    \end{figure}

The kinetic nature of the discharge process at the first microsecond represents itself an essential advantage of the model \cite{1} with respect to \cite{3}. The model \cite{1} also allows one to calculate numerically the evolution of various physical characteristics during the discharge process. Fig.3 presents the simulated graphs for the spark gap channel width $r_z(t)$, the electron number density $n_e(t)$, the mobility $\mu _e(t)$, the conductivity $\sigma (t)$, the collision frequency $\nu _c$, the ionization degree $n_e(t)/n_{mol}$, the magnetic field $H(t)$ in the discharge channel, the channel inductance $L(t)$ and the electron drift velocity $\omega _e(t)=\mu _e(t)E(t)$. Solid lines correspond to simulations with $r_{z1}(t)=r_z(t)$, where $D(t)$ is determined by Eq. (7), dotted ones to that with $r_z=const=10$ microns (we simulated also larger values of $r_z$, however they almost did not affect the time dependencies of the other physical quantities) and dashed with $r_{z2}(t)=r_z(t)$, where $D(t)$ is determined by Eq. (8).

As seen from Fig.3 the electron mobility, the elastic collision frequency of a free electron with the air molecules, the channel inductance, the electron drift velocity as well as the ohmic resistance are almost independent of $r_z(t)$, although the electron density and conductivity are very sensitive to it. Namely the graphs $n_e(t)$ and $\sigma (t)$ show that at $r_z=const$ the density and the conductivity reach their maxima at the current maximum, while for $r_z(t)$ even slightly varying with respect to $r_z=const=10$ microns their maxima shift in time to the beginning of the process, that is more consistent with the actual physical picture of the process.

\section{Conclusion}

We have simulated the residual ohmic resistance in the channel of an air spark gap using two different physical models and estimated its value as 0.2 - 0.4 Ohm in the case of the simple electrical circuit. The simulated values agree well between themselves and with the experimental data. We also show that the time dependent width of the discharge channel allows one to describe the physical processes in a spark gap much more correctly than the constant one.

\bigskip

\section{Acknowledgements} We would like to thank Prof. V.G. Baryshevsky for support given to this work and the experimental group of Institute for Nuclear Problems including A.A. Gurinovich, N.A. Belous, I.I. Vasiliev and S.I. Agafonov for given experimental data.

\end{document}